\documentclass{aa}
\usepackage{graphicx}
\begin{document}
\title{A multiwavelength study of the remnant of nova GK Persei}
\author{G. C. Anupama \inst{1} \and N. G. Kantharia \inst{2}}
\offprints{G.C. Anupama}
\institute{Indian Institute of Astrophysics, II Block Koramangala, Bangalore 
560 034, India\\
\email{gca@iiap.res.in}
\and
National Centre for Radio Astrophysics, Tata Institute of Fundamental Research,
Post Bag 3, Ganeshkhind, Pune 411 007, India
\email{ngk@ncra.tifr.res.in}}

\date{Received: 16/11/2004; accepted: 25/01/2005}

\abstract{We present new observations of the nebular remnant of the old 
nova GK Persei 1901, in the optical using the Himalayan Chandra Telescope (HCT)
and at low radio frequencies using the Giant Metrewave Radio Telescope (GMRT). 
The previous such study was by Seaquist et al (1989, hereafter S89).
The dimensions of the shell in the optical emission lines of [NII], [OIII]
and [OII] are $108\times 94$~arcsec$^2$, $104\times 94$~arcsec$^2$ and
$99\times 87$~arcsec$^2$, respectively. The evolution of the nova remnant 
indicates shock interaction with the ambient medium, especially in the 
southwest quadrant. Application of a simple model for the shock
and its evolution to determine the time dependence of the radius of the shell
in the southwest quadrant indicates that the shell is now expanding into
an ambient medium that has a density of 0.12~cm$^{-3}$, compared to the density
of the ambient medium of 0.8~cm$^{-3}$ ahead of the shock in 1987.
There are indications of a recent interaction of the nova remnant with the
ambient medium in the northeast quadrant also. There is a distinct flattening
of the shell, as well as an increase in the number and brightness of
the knots in the region. The brightest optical knots in this region are also
detected in the radio images.

The nova remnant of GK Per is detected at all the observed radio frequencies
and is of similar extent as the optical remnant.
Putting together our radio observations with VLA archival data on GK Per from 1997, we
obtain three interesting results: 1. The spectrum 
above 1.4 GHz follows a power law with an index $-0.7$ ($S\propto \nu^{\alpha}$)
and below 1.4 GHz follows a power law with an index $\sim -0.85$.  This could be due to
the presence of at least two populations of electrons dominating the global
emission at different frequencies. 
2. We record an annual secular decrease of 2.1\% in the flux density of the nova
remnant at 1.4 and 4.9 GHz between 1984 and 1997 which has left the spectral
index unchanged at $-0.7$.  No such decrease is observed in the flux densities
below 1 GHz.
3. We record an increase in the flux density 
at 0.33 GHz compared to the previous estimate in 1987.  
Finally from the above, we conclude that the remnant of nova GK Per is 
similar to supernova remnants and in particular, to the young supernova remnant
Cas~A. 
 
\keywords{novae -- shells -- individual: GK Persei}}

\titlerunning{The remnant of nova GK Persei}
\authorrunning{Anupama \& Kantharia}
\maketitle

\section{Introduction}

Cataclysmic variables (CVs) are interacting binary systems with a white dwarf
primary accreting material from its Roche-lobe filling main-sequence companion
(Warner \cite{w95}). These systems have short orbital periods of the order of a few
hours. Classical novae are a subset of CVs which undergo outbursts on
the surface of the white dwarf due to thermonuclear runaway in the accretion
material. The outbursts result in the ejection of $\sim 10^{-4}$~M$_\odot$  
of material at velocities up to several thousand kilometers per second.
GK Persei is a classical nova system that had an outburst in 1901, and is
unique in many respects. It was the first bright nova of the twentieth century
and is now classified as a fast `neon' nova. The nova binary consists of a
magnetic white dwarf with a surface magnetic field of $5\times 10^5$~G
(Bianchini \& Sabbadin \cite{bs83}) and an evolved late-type (K2IV) companion. The
orbital period of 1.904 days is the longest known for a classical nova system.
In contrast with most old novae, this system exhibits dwarf nova-like outbursts
of $\sim 3$~mag.

A month after the 1901 outburst, rapidly expanding nebulosities were detected 
in the vicinity of the nova (e.g.\ Ritchey \cite{r01}; Perrine \cite{p02}). This first 
recorded `light echo' was explained in terms of reflection from dust grains
lying in a plane crossing the line of sight to the nova (Couderc \cite{c39}). The
expanding ejecta from the nova outburst was first discovered in 1916 and has been observed
intermittently since then in the broad bands as well as in emission lines.
Seaquist et al.\ (\cite{s89}; hereinafter S89) present a detailed evolution of 
the
optical nebulosity until 1988. The sequence of the optical images indicate
the ejecta are being decelerated, particularly in the southwest quadrant, by
interaction with the ambient medium. Post S89, the optical evolution of the
shell has been studied by Anupama \& Prabhu (\cite{ap93}), Slavin et al.\ 
(\cite{sl95}) and Lawrence et al.\ (\cite{la95}).

The confirmation of the strong interaction in the southwest quadrant comes from
the detection of a ridge of non-thermal, synchrotron, emission in the radio
coincident with the flattening in the optical images in the southwest quadrant
(S89). ROSAT HRI images and the CHANDRA images in the X-rays (Balman \& 
\"Ogelman \cite{bo99}, Balman \cite{bl02}) indicate that the nebula is 
brightest in the southwest quadrant
and towards west with a lumpy morphology and resembles the radio shell. The
X-ray shell is predicted to be in a transition to the Sedov phase. The nova
ejecta of GK Per resembles a young supernova remnant.

A search for the ambient medium (Bode et al. \cite{b87}, S89) resulted in 
the detection of an extended emission in the far-infrared. This extended 
emission runs roughly northwest to southeast through the position of the nova 
and has a total extent of 30~arcmin. It is double peaked with the nova lying 
on a saddle point between the peaks. A similar extended emission is seen in HI 
and CO emission also (S89, Scott et al. \cite{s94}). A re-analysis of the high
resolution IRAS images by Dougherty et al. (\cite{d96}) indicates that the 
extended far infrared emission is due to dust at $T=23\pm 1$~K with $M_d=0.04\,
\rm{M}_\odot$, while the HI mass in the same region is 1~M$_\odot$.

Faint optical bipolar nebula, extending well beyond the current nova remnant was
first detected by Tweedy (\cite{t95}) and subsequently by Bode et al.\ 
(\cite{be04}). A detailed study of the images of the nebulosity in H$\alpha$ 
and [OIII] emission by Bode et al.\ (\cite{be04}) indicate it to be correlated 
with the light echoes detected
during the 1901 outburst. The brightest region of this large scale optical
nebulosity is also coincident with the longest lived of the light echoes of 
1902. The encountering the the expanding nova ejecta with the high density
regions of this bipolar nebula leads to shocks and particle acceleration that
are seen as radio and X-ray emission, and also in the clumpy nature and
deceleration of the nova remnant in the optical.

Bode et al. (\cite{b87}) first suggested that the IRAS emission might be due to
material ejected in a previous phase of the evolution of the central binary
and could be a fossil planetary nebula. Dougherty et al.\ (\cite{d96}) and
Bode et al.\ (\cite{be04}) later
suggested the origin of the material to be a `born again' AGB phase of the
binary as the white dwarf accreted material at a very high rate from the
secondary star creating a common envelope which was then ejected. This
evolutionary model is consistent with the current mass of the secondary which
is low for its luminosity, the $\sim 1$~M$_\odot$ material present in 
the fossil material, and the detection of the extended bipolar optical nebula.
An estimate of the proper motion of the central system together with the radial 
velocity of the system indicates a space velocity of $45\pm4$~km~s$^{-1}$
(Bode et al.\ \cite{be04}). 

We present in this paper a study of the nebular remnant in low-frequency radio
emission as well as in the optical. The evolution of the nova remnant since the
detailed work by S89 is presented. We also present a study of the environment
in HI. 
We follow Mclaughlin (\cite{m60}) in 
assuming that GK Per is located at a distance of 470 pc.  At this distance,
$1^\prime$ corresponds to 0.14 pc.

\section{Observations and Data Analysis}
\subsection{Optical}

Optical CCD images of the nebular shell around GK Per were obtained through 
Bessell UBVR, H$\alpha$+[NII] (100 \AA\ bandpass) and [OIII] (100 \AA\ 
bandpass) 
filters on 2 January and 24 November 2003 using the HFOSC instrument 
on the 2-m Himalayan Chandra Telescope (HCT). The total field of view is 
$10^\prime \times 10^\prime$ with an image scale of 0.297 arcsec/pixel. More 
details on the telescope and the instrument may be obtained from 
http://www.iiap.ernet.in/$\sim$iao.

The details of the observations are given in Table \ref{tab1}. The seeing was
poor on both occassions, and ranged between $2^{\prime\prime} - 
2.5^{\prime\prime}$. All images 
were bias subtracted and flat-field corrected in the standard manner using the 
various tasks under the IRAF package. 

\begin{table}
\caption{Optical observations}
\begin{tabular}{lcc}
\hline\hline
Date & Filter & Exposure time\\
(2003)     &        &   (seconds)\\
\hline
January 2 & $U$ & 2400\\
               & $B$ & 540\\
               & $V$ & 180\\
               & $R$ & 120\\
               & $I$ & 180\\
               & H$\alpha$ + [N\,II] & 2400\\
               & [O\,III] & 3900\\
November 24 & $U$ & 4200\\
            & $B$ & 240\\
            & $V$ & 180\\
            & $R$ & 65\\
            & $I$ & 20\\
            & H$\alpha$ + [N\,II] & 1800\\
            & [O\,III] & 4200\\ 
\hline
\end{tabular}
\label{tab1}
\end{table}

\subsection{Radio }

GK Per was observed in the radio continuum at 0.33 GHz, 0.61 GHz and 1.28 GHz 
and in 21 cm HI using the Giant Metrewave Radio Telescope (GMRT). All 
observations, except in the 1.28 GHz, were made within a couple of months
(August-October 2002). The 1.28 GHz radio continuum observations were made a 
year later in October 2003.
The observations started and ended with a run on an amplitude calibrator (3C147, 3C286) which
also doubled as bandpass calibrators for the radio continuum observations.
3C287 was used as the bandpass calibrator for the 21\,cm HI observations.
The on-source runs were interspersed with short runs
on a phase calibrator (0432+416).  We obtained an average of 5-6 hours
on-source data for all the bands.  The observational details are summarized in Table \ref{tab2}.

\begin{table*}
\caption{Details of the Radio Observations}
\begin{tabular}{lcccc}
\hline \hline
Detail  & 0.33 GHz & 0.61 GHz  & 1.28 GHz  & 21 cm \\
\hline
Date of Observation & 20/8/2002 & 5/9/2002 & 16/10/2003 & 25/9/2002 \\
Baseband Bandwidth (MHz) & 16  & 16 & 16 & 1 \\
Flux density of phase cal 0432+416 (Jy) & $17.15\pm0.08$ & $14.3\pm0.18$ & $10.9\pm 0.13$  & $10.5\pm1.2$\\
Primary beam size & $108'$ & $54'$ & $26'$ & $24'$ \\
Synthesized beam$^1$ &$13''\times11''$ & $14.4''\times 11''$ & $25''\times12.5''$ &  $37''\times28''$\\
Continuum/line rms (mJy/beam) &0.7 &0.35 &0.14 & 2\\
\hline
\\
\multicolumn{5}{l}{$^1$ This is the synthesized beam of the images presented in
the paper and, in most cases, }\\ 
\multicolumn{5}{l}{\,\, is larger than the best achievable.}\\
\end{tabular}
\label{tab2}
\end{table*}

The data were converted to standard FITS format and imported to NRAO AIPS.
The GMRT always works in the spectral line mode giving 128 channels for each 
sideband and each polarization. Hence, the GMRT data has to be first gain 
calibrated, then bandpass calibrated and then the channels averaged to obtain 
the continuum database.
Thus the analysis strategy in AIPS, generally followed for
GMRT continuum data, is that a single channel data on the amplitude and bandpass
calibrators is initially examined for bad data, edited and then gain calibrated.
The bandpass calibrator is then used to generate the bandpass gain solutions.
The bandpass calibration is then applied to all the data and line-free/RFI-free
channels averaged to generate a continuum uv data base. The continuum data on 
the phase calibrator is then used to gain calibrate the target field. The 
calibrated target field is then imaged. All the fields have been corrected for 
the primary beam gain variation.

Since wide-field effects start dominating at the lower frequencies, we imaged
the 0.33 GHz primary beam using 25 facets and the 0.61 GHz primary beam
using 9 facets. The 1.28 GHz and 21\,cm data were imaged
using simple 2-D FFT algorithms in AIPS. 

The HI 21\,cm line data was treated slightly differently than the radio 
continuum bands. The data was bandpass calibrated using 3C287. A data cube was 
then generated from this calibrated database on GK Per by using a uvtaper of
20 k$\lambda$.  The channel separation was 1.6 km\,s$^{-1}$ and the rms noise on
the channel images was 2.3 mJy/beam.

We have also analysed the 1.4 GHz and 4.9 GHz VLA data on GK Per obtained
in 1997 and available in the NRAO Data Archives.

\begin{figure*}
\centering
\caption{The nova remnant in (a) H$\alpha$+[N II], (b) [O III], and (c) [O II].
All images are $2.9^\prime \times 3.2^\prime$.
(d) The extended bipolar nebula in H$\alpha$. Note the 'jet' feature in the
northeast quadrant. The image is $10^\prime \times 10^\prime$. All images are 
with north up and east to the left. The scale bar is $30^{\prime\prime}$ in 
length.}
\label{fig1}
\end{figure*}

\section{Results}

\subsection{Optical images}

The nebular remnant of the 1901 nova outburst is clearly detected in the
H$\alpha$+[NII] and [OIII] images. The shell is also clearly seen in the $U$
band images, and could possibly be due [OII] 3727~\AA\ emission.

Fig. \ref{fig1}a shows the remnant in the H$\alpha$+[NII]. The images
of the remnant presented by Slavin et al.\ (\cite{sl95}) and Lawrence et al.\
(\cite{la95}) clearly indicate that
the H$\alpha$+[NII] image presented here is predominantly due to emission from
[NII]. The shell is boxy and asymmetric, with a flattening in the northeast
and southwest quadrants (see Figure \ref{fig1}). The major axis of the shell 
lies
approximately southeast to northwest, with a major to minor axis ratio
of 1.15. The shell is inhomogeneous and consists of blobs of emission of
varying size and surface brightness, with the bulk of the emission arising
from the southern region. The remnant appears to be three quarters
of a rectangle, with a depletion of emission knots in the eastern edge. The
shell extends to only 45~arcsec in the east, while it extends to 53~arcsec
in the west. The shell appears more symmetric along the north-south direction,
with the radius of the shell being approximately 49~arcsec in both directions.
The dimension of the shell along the major and minor axes is
$108 \times 94$~arcsec$^2$, while it was $103\times 90$~arcsec$^2$ in 1993
(Slavin et al.\ \cite{sl95}).

Fig. \ref{fig1}b shows the shell in [OIII] emission. The general morphology of the
shell is similar to [NII]. The flattening in the northeast quadrant is present
in the [OIII] image also. The dimension of the shell is
$104\times 94$~arcsec$^2$, with a major axis to minor axis ratio of 1.1.
In contrast with the [NII] emission, the emission
knots are more evenly distributed in [OIII]. The eastern hole is also not very
prominent. On the other hand the shell is more extended in the east compared
to its extent in the west.

Fig. \ref{fig1}c shows the shell in the [OII] emission. The general morphology is
very similar to the [NII] emission. The dimension of the shell is
$99\times 87$~arcsec$^2$ and the axes ratio is 1.14, similar to the
[NII] shell. The emission `hole' in the east is very prominent, and as in the
case of [NII], the bulk of the emission is in the south-southwest region.

Faint optical nebulosity extending well beyond the current nova ejecta was
first detected by Tweedy (1995) in both H$\alpha$ and [OIII] emission.
Bode et al.\ (\cite{be04}) have also detected this nebulosity in both 
emissions. Our H$\alpha$+[NII] and [OIII] images also show this extended 
nebulosity. The nebulosity in H$\alpha$ is shown in Fig. \ref{fig1}d.

The nebulosity has an ``hourglass'' shape that is
slightly tilted, and also flattened towards the southwest. The brightest
regions are the flattened portion in the southwest, and a `jet-like feature'
in the northeast. The [OIII] emission lies inwards of the H$\alpha$ emission,
except for the `jet-feature' that is coincident in both emissions. Comparing
with the IRAS 100$\mu$m images and the light echo images seen in 1902, Bode
et al.\ (\cite{be04}) find that the southwest flattening in the [OIII] 
coincides with the
infrared emission, while in the H$\alpha$, it coincides with the longest lived
of the light echoes of 1902. Bode et al.\ (\cite{be04}) suggest the flattening 
is caused by
a deceleration of the nebula due to interaction with the interstellar medium
in the direction of the motion of the system.

\begin{figure*}
\centering
\caption{GMRT naturally weighted 0.33 GHz (left), 0.61 GHz (center) and 
1.28 GHz (right) images of the radio emission from GK Persei remnant.  
The angular resolution is $13''\times11''$ at a PA$=61^{\circ}.2$ and the rms 
noise on the 0.33 GHz image is 0.7 mJy/beam. The grey scale ranges from 1.4 
($2\sigma$) to 5 mJy/beam. The angular resolution is $21.9'' \times 13.0''$
at a PA$=69.9^\circ$ and the rms noise on the 0.61 GHz image is 0.35 mJy/beam. 
The grey scale ranges from ($2\sigma$) to 2 mJy/beam.
The angular resolution is $25.04'' \times 12.52''$  at a PA$=-25.5^{\circ}$ and 
the rms noise on the 1.28 GHz image is 0.14 mJy/beam. The grey scale ranges 
from $2\sigma$ to 3 mJy/beam. In all images, the contours are plotted starting 
from $3\sigma$ and increased in steps of $\sqrt(2)$. Note that we do not detect 
the extended low brightness emission in the northeast at 0.61 GHz and 1.28 GHz. 
The cross marks the position of the nova.}
\label{fig2}
\end{figure*}

\subsection{Radio}

The naturally weighted images of the nova remnant at 0.33 GHz, 0.61 GHz 
and 1.28 GHz are shown in Figure \ref{fig2}. A uvrange of 20$k\lambda$ was 
used at 1.28 GHz so as to match the 0.33 GHz uv-cutoff. While the grey scale 
in the figure has been plotted down to $2\sigma$, the contours start from
$3\sigma$. The southwest ridge of emission is seen at all frequencies. 
While the locations of the peak emission at the 0.33 GHz and 0.61 GHz 
are not too different, the location of the peak emission at 1.28 GHz is quite
different. Moreover, the 0.61 GHz image shows features evident
in the images at the other two frequencies.

We would like to point out that there are three 
strong sources (we refer to these as source A, B following S89 and source C) 
which appear in the primary beam (see Table \ref{tab2}) at all the 
observed frequencies and limit the dynamic range of our images.
 
\begin{table}
\caption{Flux density of GK Per at the three observed frequencies.
The spectral index is calculated for the frequency in the same row and the next
frequency. The spectral index $\alpha$ ($S_\nu \propto \nu^\alpha$) in the 
first row is between 0.33 and 0.61 GHz and in the third row is between 0.33 and 
1.28 GHz. }
\begin{tabular}{cccc}
\hline \hline
Frequency & Flux density & Spectral index $\alpha$  & S89 \\
GHz & mJy &  & mJy \\
\hline
0.33 & $36.8 \pm 4$ & $-0.85 (+0.2,-0.15)$  & 20 (+10,-5) \\
0.61 & $21.8 \pm 0.5$ &$-0.87 (+0.08,-0.11)$ & $23\pm 3$ \\
1.28 & $11.4 \pm 0.5$ & $ -0.86 (\pm 0.11)$ & - \\
\hline
\end{tabular}
\label{tab3}
\end{table}

The flux density of the remnant of GK Per estimated at the three radio 
frequencies and the spectral index implied by these are listed in Table 
\ref{tab3}. The spectral index between 0.33 and 0.61 GHz varies from $-0.4$ to 
$-1.6$ across GK Per. The spectral index is steepest along a ridge running from 
north to southwest.  The global spectral index of the remnant between 0.33 GHz 
and 1.28 GHz is $-0.85$.  Unlike S89, whose data suggested that the radio 
spectrum of GK Per turned over around 1 GHz, we find that the spectrum is 
clearly a power law down to 0.33 GHz.
S89 estimated a spectral index of $-0.7$ between 1.4 GHz and 4.9 GHz.
Analysis of the VLA archival data from 1997 at 1.4 GHz and 4.9 GHz gives a 
similar spectral index. It thus appears that the spectral index above 1.4 GHz
is $-0.7$ and below 1.4 GHz is $-0.85$. There does appear to be
a spectral break near 1 GHz but the spectrum becomes steeper at lower
frequencies.

\subsubsection{21\,cm HI}

\begin{figure}
\resizebox{\hsize}{!}{\includegraphics{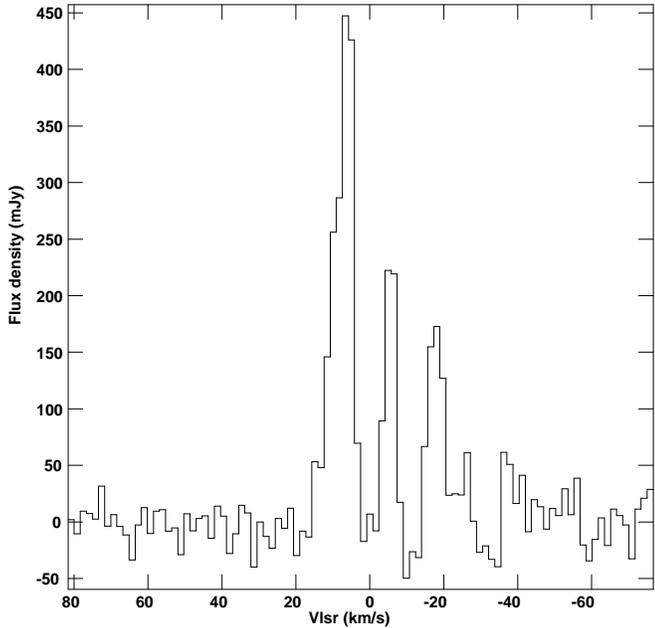}}
\caption{21cm HI emission integrated over a small region in the northwest of the nova GK Per.}
\label{fig3}
\end{figure}

21\,cm HI emission at 4.9 km\,s$^{-1}$ was detected in a $\sim 15^\prime$ 
strip running northwest to southeast with a gap in the central parts near the 
nova position. This corresponds to a linear extent of about 21 pc. 
This has been earlier reported by S89 who also found it to coincide with the IRAS
FIR emission from dust.  A spectrum integrated over a small region in the north 
of this emission is shown in Fig \ref{fig3}. The emission extends from about
20 km\,s$^{-1}$ to $-25$ km\,s$^{-1}$ and is strongest near 4.9 km\,s$^{-1}$, 
which is similar to what S89 found. The HI emission extends on both sides of 
the nova and is along the ridge of radio continuum emission from the nova 
remnant, indicating that the HI emission is related to the nova.
The HI emitting region is
large and there is considerable flux missing at low spatial frequencies
due to missing short spacings in the interferometer data.  Thus we do not
estimate the column density and mass of HI from this data.
We also detect HI emission from a region to the west of GK Per near $-5$ km\,s$^{-1}$.

\begin{table*}
\caption{HI absorption and flux densities of the sources A, B and C in the GK Per field.  
The spectral indices between two pairs of frequencies are also listed. } 
\begin{tabular}{ccccccccccc|cccc}
\hline\hline
 & RA (2000) & Dec (2000) & \multicolumn{8}{c|}{Radio Continuum} & 
\multicolumn{4}{c}{HI}\\
 & h m s & $\circ$ ' '' & S$_{330}$ & S$_{610}$ & S$_{1280}$ & S$_{1420}$ & $\alpha^{330}_{610}$ & $\alpha^{610}_{1420}$  & S89 & NVSS & \multicolumn{2}{c}{GMRT}& \multicolumn{2}
{c}{S89} \\
       & & & mJy  & mJy & mJy & mJy & & & S$_{1490}$ & S$_{1420}$ & $\tau$ &  V$_{lsr}$ & $\tau$ & V$_{lsr}$ \\
& & &  &  &  &  & & & mJy & mJy & & km\,s$^{-1}$ & & km\,s$^{-1}$   \\
\hline
& & & & & & & & & && & & &\\
A & 03 30 32.2 & $+43$ 40 02 & 614 & 369 & 192.7 &  177 & $-0.8$ & $-0.9$ & 81.1 & 179 &  0.24 & 3.0 & 0.23 & $-4.4$ \\
& & & (4) & (2) & (0.3) & (3)  && & & & & && \\
B\rlap{*} & 03 31 42.6 & $+44$ 13 10 & 848 & 689 & 249.5 & 237 & $-0.3$ & $-1.3$ & 85.3 & 335 & 0.24 &  1.9 & 0.34 & $-3.5$  \\
& & & (2) & (2) & (0.14)  & (1)  && & & & & & &\\
C & 03 31 12.4 & $+43$ 56 51 & 47 & 47 & 33.3 & 25.8 & 0 & $-0.7$ &- & 26 & 0.35 &  $-1.5$ &- &  - \\
& & & (2) & (0.7) & (0.14) & (1) & & & &  & & & &\\
& & & & & & && &  & & &\\
\hline
\multicolumn{15}{l}{\rlap{*}\ \ Source B lies close to the half power points 
of the GMRT 21\,cm primary beam and hence the values based on this}\\
\multicolumn{15}{l}{\ \ might be in error.}
\end{tabular}
\label{tab4}
\end{table*}

\begin{figure}
\resizebox{\hsize}{!}{\includegraphics{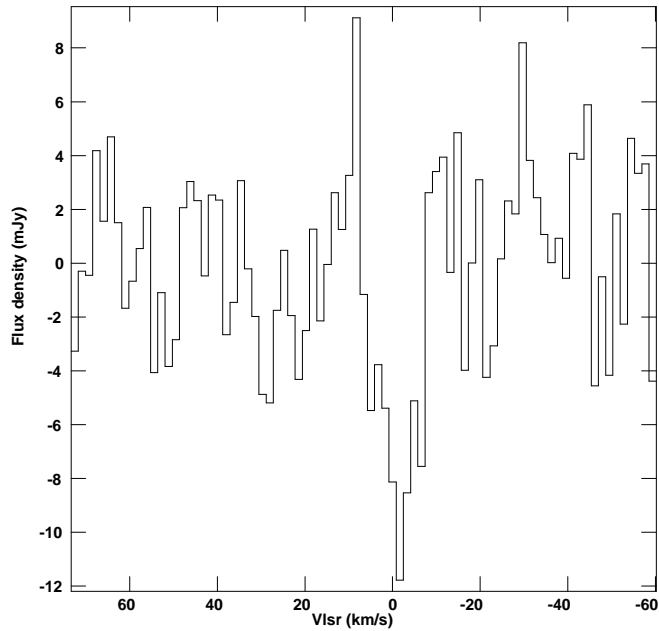}}
\caption{21cm HI absorption towards Source C located $\sim 3'$ north of GK Per.}
\label{fig4}
\end{figure}

To further examine the neutral gas near GK Per, we obtained the absorption 
spectra towards the three strong sources seen in the primary beam.  
The flux densities of these sources at the three observed frequencies
and the HI optical depth in these directions are listed in Table \ref{tab4}.
The flux density of source B at 1.4 GHz is uncertain since it lies 
close to the edge of the primary beam. The flux density of these objects that 
we estimate at 1.28 GHz and 1.4 GHz differs from those noted in S89 by a 
significant amount. 
To resolve the issue, we compared our flux densities with NVSS and find a good 
correlation as noted in Table \ref{tab4}. We, therefore, believe that the 
values 
quoted by S89 are in error, probably because they did not correct for the 
primary beam gain variation.  We detect the largest 
optical depth ($\sim 0.35$) at a velocity of $-1.5$\,km\,s$^{-1}$ towards
source C, which is about $3^\prime$ north of GK Per. The absorption spectrum 
towards this source is shown in Fig \ref{fig4}.  This extragalactic source, 
which is projected onto the extended HI nebula, is not mentioned by S89. 
Moreover, we do not detect any absorption near 4.9 km\,s$^{-1}$ in front
of source C, although the emission observed at this velocity seems to extend
to the source. This could be due to the absence of detectable absorbing gas in
front of the source at this velocity. Using the column density of
$5.4\times 10^{20}$~cm$^{-2}$, for the 5 km~s$^{-1}$ and 0 km~s$^{-1}$
features (S89), and for a limiting optical depth of $0.2$, we estimate that 
the spin temperature of the gas near 5 km~s$^{-1}$ should be $>150$~K since
we do not detect it in absorption against source C.
We do not detect significant HI emission centred near $-1.6$ km\,s$^{-1}$ where 
we detect absorption towards Source C. The spectral index of the background 
sources are also noted in Table \ref{tab4} for completion. Source A is a steep 
spectrum source with a spectral index of $\sim -0.8$, source B turns over at 
low frequencies, while source C shows a flat spectrum below 0.61 GHz. 

The local standard of rest velocities of the HI that we detect in absorption
towards sources A and B are different from what S89 have reported (see Table \ref{tab4}). 
The velocities for objects A and B vary by 7.4 and 5.4 km\,s$^{-1}$ respectively.
We think that these could be due to a systematic difference of 
about 5-6 km\,s$^{-1}$ between our results and S89 
results.  Note that the S89 data had a spectral resolution of 5 km\,s$^{-1}$, while 
our data have a resolution of 1.6 km\,s$^{-1}$. In addition, although like S89, 
we find that the extended bipolar nebula emits significantly near
5 km\,s$^{-1}$, unlike S89, we do not detect much emission near 0 km\,s$^{-1}$,
which could again be due to the systematic velocity difference.

\section{Discussion}

\subsection{Evolution of the optical remnant}

The evolution of the nova shell in the optical from 1917 to 1984 is presented
by S89. Anupama \& Prabhu (\cite{ap93}) present the images of the
shell obtained in 1990, while Slavin et al.\ (\cite{sl95}) present observations 
made in 1993. Based on a detailed study of the evolution of the remnant in the 
optical and its properties in the radio, S89 interpret the shell is 
interacting with its environment, and the flattening in the south-west is due
to deceleration caused by a shock interaction of the ejecta with its ambient
medium and conclude that in many respects that nova remnant of GK Per
behaves like a young supernova remnant. A similar explanation is provided by
Bode (\cite{b04}).

S89 also provide a simple model for the 
shock and its evolution and determine the time dependence of the radius of the
shell in the southwest quadrant. Based on both the energy conserving and
momentum conserving models for the shock interaction between the nova shell
and the ambient medium, they estimate the density of the medium ahead of the
shock to be $0.8\,\rm{cm}^{-3}$.

\begin{figure}
\resizebox{\hsize}{!}{\includegraphics{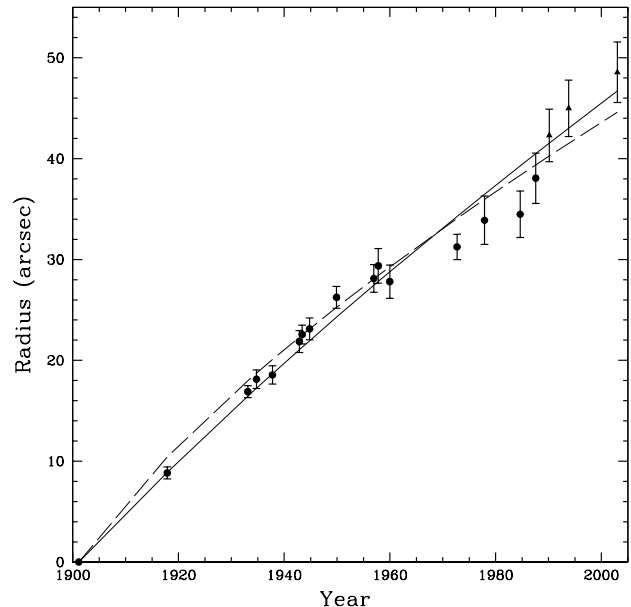}}
\caption{Time evolution of the radius of the shell in the southwest quadrant.
Filled circles refer to the radius during 1901--1988 presented in S89, while 
filled triangles refer to points later than 1988. The smooth curve corresponds
to a non-linear chisquare fit to the radius, for the energy conserving model of 
S89. The fit assumes there are no systematic differences in the estimates
of the radius.
The dotted curve represents the fit obtained using the parameters estimated by 
S89.}
\label{fig5}
\end{figure}

\begin{figure*}
\centering
\caption{0.33 GHz (left), 0.61 GHz (center) and 1.4 GHz (right) radio 
emission contours superimposed on greyscale image of the remnant in 
H$\alpha$+[NII]. The 1.4 GHz data is from the VLA Data Archives.}
\label{fig6}
\end{figure*}

Fig. \ref{fig5} shows the evolution of the radius of the southwest quadrant 
with time
for the period 1901--2003. The radii for the period 1901--1988 are obtained
from S89, at 1990 from Anupama \& Prabhu (\cite{ap93}) and at 1993 from Slavin 
et al (\cite{sl95}).
From the plot, the deceleration of the shell is evident, especially after 1950.
There is also an indication that the deceleration rate may have decreased 
since 1987. This apparent change in the deceleration rate could be a result
of systematic differences in the estimation of the radius of the shell by
various authors, e.g.\ due to different procedures adopted, seeing and
sensitivity. On the other hand, the observed change in the 
deceleration rate could be a true change, caused by density inhomogeneities 
in the ambient medium. Observations at future epochs are required to resolve 
this issue. 

A proper motion study
of the knots by Anupama \& Prabhu (\cite{ap93}) indicated the ambient density 
at the radius corresponding to the shell radius in 1990 to be lower than that 
estimated at the location corresponding to the shell radius in 1987 by S89. 
Assuming the radius estimates are not affected by systematic differences,
we fit both the energy conserving and momentum conserving shock interaction 
models of S89 to the evolution of the radius in the southwest deceleration 
zone. Using a non-linear chisquare fit (Levenberg-Marquardt algorithm: Press 
et al. \cite{p93}) for the expressions given in equations 9 and 10 in S89, we 
estimate an initial velocity ($V_0$) for the shell to be 1240 km s$^{-1}$ and 
the density of the ambient medium at $r=3.40\times 10^{17}$~cm, the current 
radius, to be 0.12 cm$^{-3}$. This value is nearly seven times lower than the 
estimate of S89 and could explain the possible decrease in the 
deceleration rate. Density inhomogenities are clearly seen in the faint 
bipolar nebula associated with GK Per and it is likely that the current 
deceleration rate would change in the future as the shell encounters 
a change in the density of the ambient medium.
The value of $V_0$ estimated here is lower than the value of 1700 km s$^{-1}$
estimated by S89, but similar to the expansion velocities detected during the
outburst in 1901. The estimated fit to the radius for the energy conserving
model is also plotted in Fig. \ref{fig5}. 

The model presented by S89 did not account for the asymmetry between the
northeast and the southwest quadrants observed at that time. S89 note that
there is no comparable interaction zone detected in the radio in the northeast
and also a lack of bright knots in that region. It is interesting to note 
the increase in the number of knots in [NII] compared to the 1990
images, and the distinct flattening of the shell in the northeast region. 
Further, emission is detected from this region, although at low flux levels, 
in the radio also. Figure \ref{fig6} shows the superposition of the 
0.33 GHz, 0.61 GHz and 1.4 GHz radio emission contours on the 
H$\alpha$+[NII] image. From the figure it is seen that the radio emission 
in 0.33 GHz and 1.4 GHz is coincident with some of the brightest optical 
knots in the northeast quadrant. This is clear evidence of 
interaction of the shell with the ambient medium in that direction (see also 
Lawrence et al.\ \cite{la95}; Slavin et al. \cite{sl95}). It 
appears that the interaction in the northeast direction is a recent phenomenon 
caused by the nova shell encountering a density enhancement in the ambient 
medium, and would be interesting to study its future development in both the
optical and radio. 

\subsection{Evolution of the radio remnant}

\begin{figure}
\resizebox{\hsize}{!}{\includegraphics{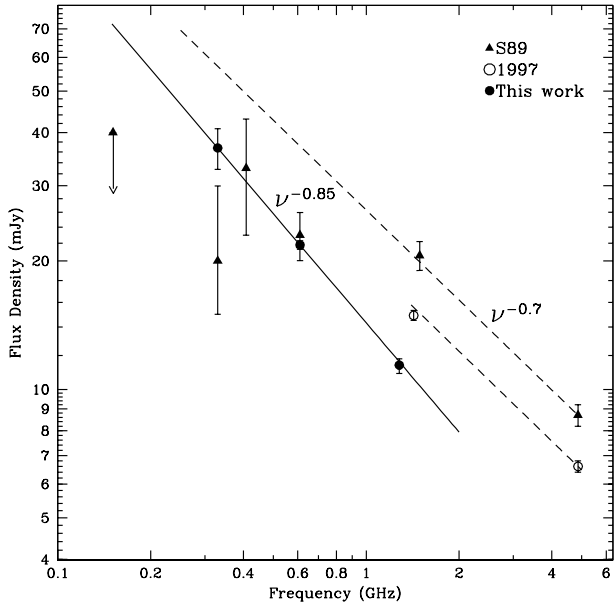}}
\caption{Radio spectrum of GK Per. Filled circles correspond to the data 
presented here, filled triangles correspond to the data presented in S89,
and open circles correspond to the flux densites in 1997.
Also shown is the power law $\nu^{-0.85}$ for the spectrum between 330 and 
1.28 GHz (continuous line), and the power law $\nu^{-0.7}$ estimated by S89 
for the spectrum between 1.4 GHz and 5 GHz (dashed lines).}
\label{fig7}
\end{figure}

The previous study of the nova remnant of GK Per in the radio by S89 indicated
a global spectral index of $-0.7$ between 1.49 GHz and 4.86 GHz, and that the 
spectrum turned over near 1 GHz. Based on this, S89 concluded that the nova
remnant is somewhat different from supernova remnants, where the turnover
occurs at much lower frequencies. The data presented here, on the other hand,
indicate that the radio spectrum of the nova remnant follows a power law
with a spectral index $\alpha \sim -0.85$ between 0.33 GHz and 1.28 GHz (Fig
\ref{fig7}). Unlike S89, we do not see a turnover around 1 GHz, and the radio 
spectrum of the GK Per shell closely resembles that of a supernova remnant. 

We record a reduced flux density at 1.28 GHz as compared to S89 by almost a 
factor of two.  We do not detect the low brightness plateau around the 
ridge of emission that S89 had reported.  It is likely that the nova shell 
has evolved over two decades.  While the reduction in the flux density in the 
1.28 GHz could probably be explained by synchrotron ageing, one needs to 
explain the absence of a turnover in the present data.  It is of interest to
note that the flux density near 0.61 GHz estimated by S89 agrees with our 
estimate within errors (see Table \ref{tab3}).  Also, the spectral index 
between 0.61 GHz and 0.408 GHz is $-0.85$ (see Fig \ref{fig7}) similar to
the spectral index estimated for the present data. The flux density near 
0.33 GHz is however discrepant; we estimate a higher flux density compared to 
S89.  It should also be noted that all the low frequency ($< 0.5$ GHz) data 
points in S89 have large error bars.  If we
assume that the flux density estimate at 0.33 GHz by S89 to be in error and more
likely similar to the present estimate, this would imply a steeper spectrum at
the lower frequencies which has not changed in two decades.  Alternatively,
it is possible that the flux at 0.33 GHz has indeed increased. 

\begin{table}
\caption{Flux density of GK Per around 1.4 GHz and 4.9 GHz.}
\begin{tabular}{ccc}
\hline \hline
Frequency & Flux density & Spectral index $\alpha$\\
GHz & mJy & \\
\hline
\multicolumn{3}{c}{S89}\\
1.490 & $20.6\pm 1.6$ & $-0.69$\\
4.860 & $8.7\pm 0.5$ & \\
\multicolumn{3}{c}{1997}\\
1.425 & $14.9\pm 0.4$ & $-0.66$\\
4.900 & $6.6\pm 0.2$ & \\
\hline
\end{tabular}
\label{tab5}
\end{table}

The evolution of the radio shell since the work of S89 is traced. The flux
densities at 1.4 GHz and 4.9 GHz in 1997 indicate that the spectral index
between 1 GHz and 5 GHz has remained the same since 1984 (S89), at $-0.7$,
while the flux densities have decreased (see Fig. \ref{fig7}). The flux
densities at these frequencies, together with the global spectral index at
1984 and 1997 are listed in Table \ref{tab5}. The flux densities imply a
secular decrease of 2.1\% per year.  Assuming a similar secular
decrease, one would expect the flux density at 1.4 GHz to be 13.1~mJy in 2003, 
while the observed flux density at 1.28 GHz is 11.4~mJy. It should be borne in 
mind that the presence of the three strong sources in the field in the 1.28 GHz 
data could affect the flux density estimate to some extent. On the other hand, 
the secular decrease in the flux density could be different from the previous secular
decrease of 2.1\% per year, but we are unable to ascertain this due to the absence of
data at higher frequencies for the same epoch. If one assumes the spectral index
remains unchanged, then, based on the observed 1.28 GHz flux density, 
we expect the flux density at 5 GHz to be 4.6 mJy in epoch 2003. 

We discuss below the most interesting aspects of the evolution of the radio
emission.
\begin{enumerate}
\item The spectrum of electrons above 1.4 GHz follows a power law
with an index $-0.7$ and below 1.4 GHz follows a power law with index
$-0.85$. There is a break near 1.4 GHz, but the change in spectral
index is negative and of a small magnitude ($0.15$) contrary to what
S89 had concluded. This we believe could be due to the presence of at least 
two populations of electrons dominating the global emission at different 
frequencies.  This is somewhat evident in Fig \ref{fig2}. The radio continuum 
emission at 0.33 GHz shows two peaks, the strongest is distinct from the peak 
of emission near 1.4 GHz. The changes in the relativistic particle 
population could be caused by density changes in the circumstellar medium with 
which the nova shell interacts. Spectral index variations have been studied in 
detail in Cas~A where different knots have been found to show different 
spectral indices (Anderson et al.\ \cite{an91}; Wright et al. \cite{w99}) and
attributed to differences in the local particle acceleration conditions.
It would be interesting to carry out a similar study of GK Per and study the
spectral index variation across the remnant.

\item The nova remnant is experiencing a secular decrease in the flux density
of about 2.1\% per year at frequencies above 1.4 GHz which has left the 
spectral index unchanged.  Although the rate of secular decrease is higher in 
GK Per, we note that an annual secular decrease of about 1\% has been 
well-studied in Cas~A, which also shows a frequency dependent decrease 
(e.g.\ Shklovskii \cite{s60}, Baars et al.\ \cite{b77}, Agafonov \cite{ag96}, 
Reichart \& Stephens \cite{rs00}). 

The secular decrease in the flux density in GK Per could be a result of 
adiabatic expansion of the remnant into the surrounding medium. 
The evolution of the optical shell indicates the shell could be in the
energy conserving (adiabatic) phase. In this case, the flux density would
vary as $S_\nu \propto r^{-2(2\alpha + 1)}$, where $r$ is the radius of the
shell and $\alpha$ is the spectral index (Shklovskii \cite{s60}). The ratio of the flux densities at 
two different epochs would be $S_2/S_1 = (r_2/r_1)^{-2(2\alpha + 1)}$. Using 
the radii estimates at 1984 and 1997 from Figure \ref{fig5}, the expected
ratio of the flux densities is 0.53. The observed ratios are 0.72 (1.4 GHz)
and 0.79 (4.9 GHz), which are similar to the expected ratio (within errors),
indicating the secular decrease can be attributed to adiabatic expansion.

\item We estimate a higher 0.33 GHz flux density compared to S89 (see Table 
\ref{tab3}). If this is indeed a true rise, then GK Per is increasingly looking 
similar to Cas A which has shown anomalous increase in the flux density at
0.038 GHz (Erickson \& Perley \cite{ep75}; Read 1977a,b; Walczowski \& Smith
\cite{ws85}). The increase in the flux density in Cas A is explained by
Chevalier et al. (\cite{cet78}) as being due to a fresh injection of 
relativistic electrons due to density variations in the vicinity of Cas~A.
\end{enumerate}

It would be interesting to model the emission from the nova remnant and arrive
at a consistent model. However, before that it is necessary to confirm 
(a) the secular decrease in GK Per, and (b) the increase in flux density
in the 0.33 GHz. Clearly, there is a need for more multifrequency, nearly 
simultaneous data in order to have a better understanding of the evolution and 
nature of the radio shell of nova GK Per.

\section{Conclusions}

We present in this paper the evolution of the nova remnant of GK Persei, in
the optical and low-frequency radio regions, since the previous detailed work 
of S89.

\begin{enumerate}

\item The shell is boxy, asymmetric and clumpy in nature.
The dimensions of the shell in the optical emission lines of [NII], [OIII]
and [OII] are $108\times 94$~arcsec$^2$, $104\times 94$~arcsec$^2$ and
$99\times 87$~arcsec$^2$, respectively. The shell has expanded since the
previous estimate in 1993 by Slavin et al. (\cite{sl95}). The evolution of the 
southwest interaction zone of the shell since 1916 indicates that the shell
is decelerating due to shock interaction of the nova ejecta with its ambient
medium (S89). Application of a simple model for the shock
and its evolution to determine the time dependence of the radius of the shell
in the southwest quadrant (S89) indicates that the shell is now expanding into
an ambient medium that has a density of 0.12~cm$^{-3}$, compared to the 
ambient medium density of 0.8~cm$^{-3}$ ahead of the shock in 1987.

\item There are indications of a recent interaction of the nova remnant with the
ambient medium in the northeast quadrant also. There is a distinct flattening 
of the shell, as well as an increase in the number and brightness of 
the knots in the region. The brightest optical knots in this region are also 
detected in the radio images.

\item The spectrum of electrons above 1.4 GHz follows a power law with an 
index $-0.7$ 
and below 1.4 GHz follows a power law with index $-0.85$. This could be due to 
the presence of at least two populations of electrons dominating the global 
emission at different frequencies.

The flux densities at the 1.4 GHz and 4.9 GHz estimated by S89 indicate a
spectral index of $-0.7$ over that frequency range. The flux densities of
the shell at the same frequencies in 1997 indicate no change in the spectral
index, while the flux densities have decreased, indicating a secular decrease
of 2.1\% per year. We are unable to comment on the present evolution of the 
nova remnant at these frequencies due to lack of observations.

The flux density that is estimated here at 0.33 GHz is much higher compared to 
S89.

It is suggested that the observed evolution of the nova remnant in the radio
is probably due to changes in the relativistic particle population caused by
density changes in the ambient medium with which the shell interacts. The
evolution of the GK Per remnant appears to be similar to that of young
supernova remnants, in particular, the young supernova remnant Cas~A.

\item The optical images also clearly show the extended bipolar nebula 
associated with mass loss processes that occurred during the evolution of the 
nova binary. This nebulosity has an hourglass
shape that is flattened towards the southwest. The brightest regions of the 
nebula are the flattened portion in the southwest and a jet feature in the
northeast. The [OIII] emission lies inwards of the H$\alpha$ emission, except
for the jet feature that is coincident in both emissions.

\item We also detect the extended HI bipolar emission feature associated with 
GK Per, 
in our 21 cm images. The emission is detected at 4.9 km~s$^{-1}$, and extends
over $\sim 15$~arcmin, which corresponds to a cloud of linear dimensions of
21 parsec at the distance of GK Per. 
\end{enumerate}

\begin{acknowledgements}
We thank the staff of IAO, Hanle and CREST, Hosakote, for their support during
obervations. The facilities at IAO and CREST are operated by the Indian 
Institute of Astrophysics, Bangalore. We thank the staff of the 
GMRT that made these observations possible. GMRT is run by the 
National Centre for Radio Astrophysics of the Tata Institute of Fundamental 
Research. This work has made use of The NRAO Data Archives. The National Radio 
Astronomy Observatory is a facility of the National Science Foundation (U.S.A.)
operated under cooperative agreement by Associated Universities, Inc.
GCA thanks K.E. Rangarajan for his help with the non-linear chisquare fit.
Discussions with A. Mangalam and D. Bhattacharyya are acknowledged.
We thank the referee for encouraging and very useful comments.
\end{acknowledgements}

\end{document}